\documentclass[amssymb,aps,twocolumn,nofootinbib]{revtex4}
\usepackage[]{graphicx}
\usepackage[]{amsmath}
\usepackage{hyperref}

\def\kb#1#2{| #1 \rangle\!\langle #2 |}

\def\cC{\mathcal{C}}

\def\cH{\mathcal{H}}

\def\cN{\mathcal{N}}

\def\Tr{\mathrm{Tr}}

\newtheorem{theorem}{Theorem}
\newtheorem{lemma}{Lemma}
\newtheorem{definition}{Definition}
\def\eq#1{Eq.~\eqref{eq:#1}}

\def\eq#1{Eq.~\eqref{eq:#1}}

\begin{document}

\title{Quantum Markov Networks and Commuting Hamiltonians}
\author{Winton Brown}
\email{Winton.Brown@USherbrooke.ca}
\affiliation{D\'epartement de Physique, Universit\'e de Sherbrooke, Qu\'ebec, Canada, J1K 2R1}
\author{David Poulin}
\email{David.Poulin@USherbrooke.ca}
\affiliation{D\'epartement de Physique, Universit\'e de Sherbrooke, Qu\'ebec, Canada, J1K 2R1}

\date{\today}

\begin{abstract}
Quantum Markov networks are a generalization of quantum Markov chains to arbitrary graphs. They provide a powerful classification of correlations in quantum many-body systems---complementing the area law at finite temperature---and are therefore useful to understand the powers and limitations of certain classes of simulation algorithms. Here, we extend the characterization of quantum Markov networks \cite{LP08a,PH11a}  and in particular prove the equivalence of positive quantum Markov networks and Gibbs states of Hamiltonians that are the sum of local commuting terms on graphs containing no triangles. For more general graphs we demonstrate the equivalence between quantum Markov networks and Gibbs states of a class of Hamiltonians of intermediate complexity between commuting and general local Hamiltonians.

\end{abstract}


\maketitle

\section{Introduction}

A Markov network is an object used to represent the correlation structure of a probability distribution over a very large number of random variables. In this representation, random variables are located at the vertices of a graph and the edges encode correlations between neighboring vertices. While distant vertices can be correlated, their correlations are mediated by intermediate random variables located along the paths connecting the two vertices; distant variables are conditionally independent given the variables separating them. Since a sequence of conditionally independent variables defines a Markov chain, Markov networks are their natural generalization from chains to arbitrary graphs \cite{Lau96a,Mac03a,Nea90a,Nea04a}.

An important motivation for the characterization of Markov networks stems from the existence of powerful algorithms and heuristics to solve or approximate the solution to statistical inference problems on these graphical models, see, e.g., \cite{Yed01a,YFW02a,AM00a} and references therein. The archetypical inference problem consists in determining the marginal probability of a given random variable after having observed some of the other random variables. Applications of such inference problems range from medical diagnostics \cite{WKAS78a} to digital communication \cite{MMC98a}, and from computer vision \cite{LE94a} to gene expression analysis \cite{JNBV11a}.

Another application, which will be the main focus in this article, is statistical physics. In this setting, particles (e.g. Ising spins) are subjected to local interactions on a regular lattice or a more general graph, and we are interested in computing some correlation function at a given temperature. The thermal equilibrium state of the system is a Gibbs probability distribution, which is a function of the interaction between particles. A powerful characterization of such graphical models, due to Hammersley and Clifford \cite{HC71a}, shows that the set of Gibbs distributions arising from local interactions on a given graph coincides with the set of Markov networks on that graph.

\begin{figure}
\includegraphics[width=4cm]{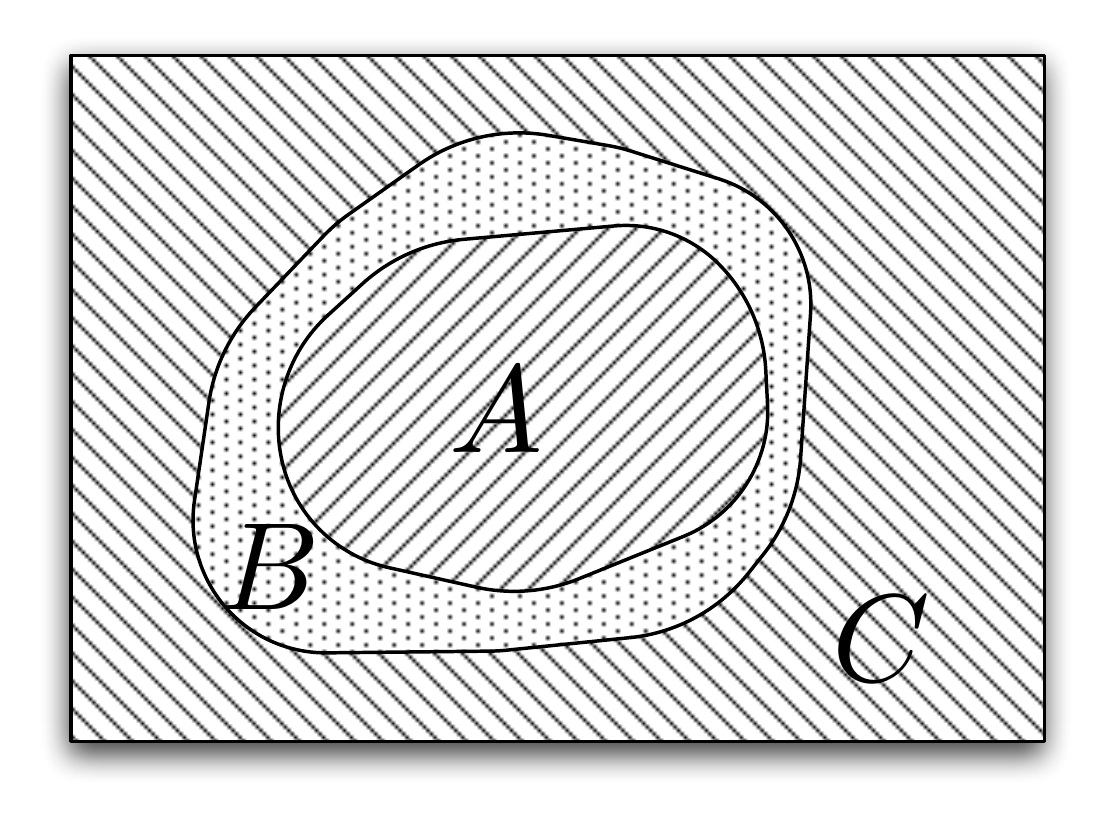}
\caption{The Markov condition demands that the mutual information between region $A$ and the rest of the system $C$, condition on its boundary $B$, be zero. More formally, this condition $I(A:C|B) = 0$ must be satisfied whenever $A$, $B$, and $C$ are disjoint regions such that all paths from $A$ to $C$ go through $B$.}
\label{fig:ABC}
\end{figure}

In this article, we study quantum mechanical version of this characterization. Quantum Markov networks, introduced in \cite{LP08a}, can be defined similarly to classical ones, in terms of conditional independence. Informally, we say that a many-body quantum state is a Markov network when the mutual information between a region $A$ and the rest of the system, conditioned on the boundary of region $A$, vanishes, see Fig.~\ref{fig:ABC}. Our general goal is to understand the relation between such states and quantum Gibbs states. The non-commutativity of quantum operators is the main obstacle to a direct generalization of the classical Hammersley-Clifford Theorem.

It was previously known \cite{LP08a} that quantum Markov networks can be expressed as Gibbs states of local Hamiltonians, but that the converse is not generally true.  While the converse holds whenever the terms in the Hamiltonian mutually commute---i.e., local commuting Hamiltonians, as we call them for short, give rise to Gibbs states that are Markov networks---it was unknown whether this commutation condition is necessary.  That is, whether there is an equivalence between positive quantum Markov networks and Gibbs states of commuting Hamiltonians.  We show, as our main result,  (Theorem~\ref{thm:2cliques}), that there exists such an equivalence between local commuting Hamiltonians and quantum Markov networks on graphs that contain no triangular cells, extending a result previously derived for cycle-free graphs \cite{PH11a}.  We also construct an example of a positive quantum Markov network on a triangular lattice, which is not a Gibbs state of a commuting Hamiltonian.

The motivation for our work stems from several sources. Quantum many-body systems are believed to obey an entanglement area law at zero temperature: in the ground state of a gapped interacting quantum system, the entanglement between a region $A$ and the rest of the system should only increase proportionally to the size of the boundary of $A$, not its volume. The area law reflects the fact that the quantum correlations in the system are short ranged. At finite temperature, it has been shown \cite{WVHC08a} that the mutual information between region $A$ and the rest of the system obeys an area law. This finite-temperature characterization does not describe entanglement, however, since both classical and quantum correlations are picked up by mutual information. Conditional mutual information could be an important tool to study entanglement area laws at finite temperature because its vanishing demonstrates that quantum correlations are short range, while classical correlations are Markovian but can extend to arbitrarily long range.

Thus, the Markov condition ensures that the quantum many-body state has only finite range entanglement. In a system with topological order, the conditional mutual information would be zero for any topologically trivial region $A$, however, there must exist a non-trivial region, for instance when $A$  is a ribbon wrapping around a torus, such that the mutual information between $A$ and the rest of the system, conditioned on the boundary of $A$, is equal to the topological entanglement entropy of the system \cite{LW05a,KP06a}. The state obtained from the uniform mixture of all ground states of a topologically ordered system, such as Kitaev's toric code \cite{Kit03a}, can form a Markov network however. 

Another goal of this line of research is to leverage the power of the various algorithms and heuristics that have been developed in the classical setting to the quantum realm. For instance, belief propagation can be used to solve statistical inference problems involving many correlated random variables. It is exact when the underlying statistical model is a Markov network on a cycle-free graph, and works remarkably well in other contexts. Belief propagation and its variational dual have been generalized to the quantum setting \cite{Has07b,LSS08a,LP08a,PH11a}. While numerical experiments have shown that they can provide reliable approximations, the present work could help understanding their validity more rigorously.

Lastly, our work is motivated by questions in quantum Hamiltonian complexity. One central question in this field of research is to determine the computational complexity of estimating the ground state energy of a quantum many-body system, or equivalently its free energy at finite temperature, for which belief propagation can be used. A special case of this question concerns local commuting Hamiltonians. Although at first glance commuting Hamiltonians seem to be entirely classical, they can exhibit important quantum phenomena such as topological order \cite{Kit03a}. It remains unclear at this stage whether the ground state of commuting Hamiltonians embody the same complexity as the ground state of arbitrary local Hamiltonians. This problem is the subject of intensive studies, see e.g.  \cite{BV05a,S11a,H12a}, and some of the questions addressed in this article are directly related to it. In particular, quantum Markov networks seem to offer an intermediate structure between commuting and general local Hamiltonians.

\section{Quantum Markov Networks}

\subsection{Quantum Markov chains}
\label{sec:chain}

Before defining quantum Markov networks, we review the more familiar concept of quantum Markov chains. Throughout this article, we consider a finite dimensional system composed of $N$ particles labeled $a=1,2,\ldots, N$, each associated to a finite-dimensional Hilbert space $\cH_a$. The Hilbert space of the entire system is $\cH = \bigotimes_a \cH_a$. The state of the system is a non-negative matrix $\rho : \cH \rightarrow \cH$ with trace equal to 1.

Conditional independence will play a crucial role in the definition of a Markov network. Given a density matrix $\rho_{ABC}$ on 3 disjoint collections of particles labeled $A$, $B$, and $C$, the conditional mutual information between $A$ and $C$ given $B$ is defined as
\begin{equation}
I(A:C|B) = S(AB)+S(BC) - S(ABC) - S(B)
\end{equation}
where $S(\rho) = -\Tr \rho\log\rho$ is the von Neumann entropy, and $S(X) = S(\rho_X)$ denotes the von Neumann entropy associated to the density matrix $\rho_X = \Tr_{\overline X} \rho$ of subsystem $X$, obtained by taking the partial trace over the complement $\overline X$ of $X$. We say that $A$ and $C$ are independent conditioned on $B$ if and only if  $I(A:C|B) = 0$.

Conditional independence implies that $A$-$B$-$C$ forms a quantum Markov chain, i.e. it is possible to create the state $\rho_{ABC}$ starting with the state $\rho_{AB}\otimes \kb 00_C$ and applying a physical transformation that involves the systems $B$ and $C$ only \cite{HJPW03a}.

A useful characterization of independence of $A$ and $C$, conditioned on $B$  \cite{HJPW03a}, is that the Hilbert space of subsystem $B$ decomposes as,

\begin{equation}
\mathcal{H}_B = \sum_J \mathcal{H}_{J:B\rightarrow A}\otimes \mathcal{H}_{J:B\rightarrow C}
\label{eq:split}
\end{equation}
such that $\rho_{ABC} = \sum_J p_J \sigma_{J:AB}\otimes \sigma_{J:BC}$ where $p_J$ is a probability distribution and each $\sigma_{J:AB}$ and $\sigma_{J:BC}$ are density matrices with support only on $\mathcal{H}_{J:B\rightarrow A}\otimes \mathcal{H}_A$ and $\mathcal{H}_{J:B\rightarrow C}\otimes \mathcal{H}_C$ respectively.

We find it convenient to express the above characterization as the following decomposition of the state into a product of commuting operators,
\begin{equation}
\rho_{ABC} =  \Lambda_{AB} \Lambda_{BC} \ \ {\rm with}\ \ [ \Lambda_{AB}, \Lambda_{BC}] = 0
\label{eq:chain}
\end{equation}
and the $\Lambda$'s are non-negative. In this equation, we make use of a notation where $Q_X$ is an operator that acts only on the Hilbert space of $X$ and is trivial everywhere else. It should be understood that $Q_X$ is an operator on $X$ tensor product with the identity on the complement of $X$, but we omit these identity matrices throughout to simplify the notation. When the density matrix is positive $\rho_{ABC}>0$, we can take the logarithm of \eq{chain} and arrive at
\begin{equation}
\rho_{ABC} =  e^{H_{AB}+H_{BC}} \ \ {\rm with}\ \ [ H_{AB}, H_{BC}] = 0,
\label{eq:chainH}
\end{equation}
which is the standard expression for the Gibbs state associated to the Hamiltonian $H = H_{AB} + H_{BC}$ (we rescale the Hamiltonian by $-1$/temperature throughout to lighten the notation).

The above construction generalizes to more than 3 systems. Consider a collection of $N$ quantum systems with density matrix $\rho$, and suppose that
\begin{equation}
I(1,\ldots k-1:k+1,\ldots,N|k) = 0
\end{equation}
	for all $k=2,3,\ldots,N-1$. These conditions can be understood by imagining that the particles are arranged in a chain, and that conditioned on any site $k$, the sites to the left of $k$ are independent of the sites to the right of $k$.  For each site, $\rho$ decomposes according to \eq{chainH}.  As will be shown in Sec.\ref{sec:comm}, the decompositions for subsequent sites may be iterated, arriving at the conclusion that $\rho = e^H$ where $H = \sum_k h_{k,k+1}$ with $[h_{k,k+1},h_{k',k'+1}]=0$. Thus, quantum Markov chains are Gibbs states of local commuting Hamiltonians. As we will see in Sec.~\ref{sec:converse}, the converse implication also holds and we have a complete equivalence between positive quantum Markov chains and local commuting Hamiltonians in 1D.

\subsection{Quantum Markov networks}

A quantum Markov network is a generalization of the above construction from a chain to an arbitrary graph. A graph $G$ is composed of a set of vertices $V$ and edges $E$. The particles are located on the vertices $V = \{1,2,\ldots, N\}$ of a graph $G = (V,E)$; we use the same label $a$ for a vertex and the particles located on it. Let $A$, $B$, and $C$ be three disjoint subsets of $V$. We say that $B$ shields $A$ from $C$ whenever all paths on $G$ beginning at a vertex in $A$ and ending at a vertex in $C$ pass through a vertex in $B$. For instance, the boundary of a region $A$---composed of all the vertices that share an edge with a vertex in $A$ but do not belong to $A$---always shields $A$ from the rest of the lattice, as illustrated on Fig.~\ref{fig:ABC}.

We say that the pair $(\rho,G)$, formed by a graph $G$ with $N$ vertices and a density matrix $\rho$ for the $N$ particles located on the graph $G$, is a
{\em quantum Markov network} when $I(A:C|B) = 0$ for all disjoint subsets $A,B,C \subset V$ such that $B$ shields $A$ from $C$. In other words, in a Markov network, $A$-$B$-$C$ forms a Markov chain when all paths from $A$ to $C$ go through $B$. Classical Markov networks are defined similarly, but with the Shannon entropy replacing the von Neumann entropy and a probability distribution replacing the density matrix.

In Ref.~\cite{LP08a}, it was shown that positive quantum Markov networks---those with $\rho>0$---are Gibbs states of local Hamiltonians. For a general graph $G$, a local Hamiltonian is defined to be of the form
\begin{equation}
H = \sum_{Q\in \cC} h_Q
\label{eq:localH}
\end{equation}
where $\cC$ denotes the set of {\em cliques} of the graph $G$. Recall that a clique of $G$ is a complete (fully-connected) subgraph of $G$. For instance, on a regular square lattice, the Hamiltonian would be a sum of terms acting on nearest neighbors, while on a triangular lattice it would also include three-body terms acting on the sites of a triangular cell.

This characterization leaves open the converse implication, namely under what circumstances do quantum Gibbs sates form Markov networks. This question is the focus of the following section.

\section{Quantum generalizations of the Hammersley-Clifford Theorem}

We begin by formally stating the classical characterization Theorem:

\begin{theorem}[Hammersley and Clifford \cite{HC71a}]
Let  $G = (V,E)$ be a graph and $P(V)$ be a positive probability distribution over random variables located at the vertices of $G$. The pair $(P(V),G)$ is a positive Markov network if and only if the probability $P$ can be expressed as $P(V) = \frac 1Z e^{H(V)}$ where
\begin{equation}
H(V) = \sum_{Q\in \cC} h_Q(Q)
\end{equation}
is the sum of real functions $h_Q(Q)$ of the random variables in cliques $Q$, and $Z$ is a normalization constant.
\end{theorem}
This Theorem establishes a complete equivalence between positive classical Markov networks and Gibbs distributions arising from local Hamiltoinians, and in the next sections we will partially generalize this equivalence to the quantum setting.

\subsection{Markov implies locality}

 One direction of this Theorem holds in the quantum setting:
\begin{theorem}[Leifer and Poulin \cite{LP08a}]
Let  $G = (V,E)$ be a graph and $\rho$ be a density matrix for the particles located at the vertices of $G$. If the pair $(\rho,G)$ is a positive quantum Markov network, then the state $\rho$ can be expressed as $\rho = e^H$ where
\begin{equation}
H = \sum_{Q\in \cC} h_Q
\label{eq:local}
\end{equation}
is the sum of Hermitian operators $h_Q$ on the particles located in cliques $Q$.
\label{thm:LP}
\end{theorem}
Here, we present a proof that is much simpler than the one presented in \cite{LP08a}, although similar in essence. The key idea is to make a cumulant expansion of the ``effective Hamiltonian", formally defined as $H = \log \rho$. Recall that any operator $H$ acting on $N$ particles can be uniquely decomposed into a sum
\begin{equation}
H = \sum_{X\in V} K_X
\label{eq:cumulant}
\end{equation}
where $X$ runs over all the subsets of $N$ particles. What makes this decomposition unique is that each cumulant $K_X$ obeys $\Tr_Y K_X=0$ for any $Y\subseteq X$, i.e. they are ``partial-traceless''. This property also implies that the cumulants are orthogonal with respect to the Hilbert-Schmidt inner product $\Tr (K_X K_Y) = \delta_{XY} \Tr K_X^2$, and it follows that $\Tr (HK_X) = \Tr K_X^2$.

Given these properties, we can easily prove the theorem. All that is required is to show that $K_X = 0$ whenever $X$ is not a clique. Consider such a region $X$. Then, there exists two vertices $a$ and $c$ $\in X$ that are not linked by an edge, and so the region $B = V-a-c$ shields $a$ from $c$. Repeating the arguments leading to  Eqs.~(\ref{eq:chain}, \ref{eq:chainH}), we arrive at $H = H_{aB} + H_{Bc}$, so
\begin{align*}
\Tr (H K_X) &= \Tr (H_{aB} K_X) + \Tr(H_{Bc}K_X) \\
& = \Tr (H_{aB} [\Tr_c K_X]) + \Tr(H_{Bc}[\Tr_a K_X]) = 0
\end{align*}
by the partial-tracelessness property of cumulants. This implies $\Tr K_X^2 =0$, and so $K_X = 0$ as claimed, which completes the proof.

\subsection{Locality and commutativity implies Markov}
\label{sec:converse}

Unlike the classical case, it is not true, however, that an arbitrary Gibbs state of a local Hamiltonian \eq{local} yields a positive quantum Markov network, as one can easily find examples of local Hamiltonians that do not generate Markov networks, even in 1D \cite{LP08a}. Specifically, any Hamiltonian that cannot always be decomposed into a sum of commuting terms $H_{AB} + H_{BC}$ across region $B$ that shields $A$ from $C$ generates a Gibbs state which violates condition \eq{chainH}. On the other hand, when all of the terms in the Hamiltonian commute, \eq{chainH} is always satisfied, implying that the Gibbs state of any local commuting Hamiltonian is a quantum Markov network.

\begin{theorem}
Let $G = (E,V)$ be a graph and $H = \sum_{Q\in \cC} h_Q$, $[h_Q,h_{Q'}] = 0$, be a local commuting Hamiltonian on that graph. Then $(\rho,G)$ is a positive quantum Markov network for $\rho = \frac 1Z e^H$, where $Z$ is a normalization constant.
\label{thm:commute}
\end{theorem}

To prove this theorem, we can first consider a restricted set of partitions $A$, $B$, and $C = V-A-B$ such that $B$ shields $A$ from $C$. In this case, we can easily show that $A$-$B$-$C$ forms a Markov chain from the observation that no clique $Q$ can overlap simultaneously with region $A$ and $C$, otherwise they would not be shielded by $B$. Thus, we can (non-uniquely) assign each clique to either region $AB$ or $BC$ and decompose $H = \sum_{Q\in AB} h_Q + \sum_{Q\in BC} h_Q$, two terms that obviously commute, ensuring that the Gibbs state is a Markov Network, c.f. \eq{chainH}. 

We can readily extend the result to an arbitrary choice of regions $A$ and $C$ that are shielded by $B$, by the following procedure: expand regions $A$ and $C$ to $AA'$ and $CC'$ until $AA'BCC'$ fills the entire lattice, and $B$ still shields $AA'$ from $CC'$. This can be done by recursively expanding region $A$ to include all of its neighbors that are not in $B$, and similarly for $C$. At the end of this process, any isolated islands that are not included in either the expanded $A$, the expanded $C$, or B can be  included in $A'$. Then, because $B$ shields $AA'$ from $CC'=V-AA'-B$, it follows from the argument of the previous paragraph that $AA'-B-CC'$ is a Markov chain, so $I(AA':CC'|B) = 0$. Then---using $I(AA':CC'|B) \geq I(A:C|B)$, which is a straightforward consequence of strong sub-additivity of entropy \cite{LR73c}---we obtain that $I(A:C|B)=0$, so $A-B-C$ is also a Markov chain.

We also note that the above result may be extended to non-positive density matrices that are projectors onto eigenspaces of arbitrary commuting Hamiltonians,  including for example the projector onto the code space of any local stabilizer code.  Since any such density matrix may be written as a product of the projectors onto the corresponding eigenspace of each term in the Hamiltonian, it follows from \eq{chain} that conditional independence is satisfied for any set of partitions $A$, $B$, and $C = V-A-B$ such that $B$ shields $A$ from $C$.  Conditional independence for arbitrary choice of regions $A$ and $C$ that are shielded by $B$, then follows similarly from strong sub-additivity of entropy, implying that any such density matrices are quantum  Markov networks.  

\subsection{When does the Markov condition imply commutativity?}
\label{sec:comm}

As seen in the proof of Theorem \ref{thm:commute}, a quantum Markov network can be defined by demanding that the condition  $I(A:C|B) = 0$ holds for any region $A$, $B$ and $C$ that span all vertices with $B$ shielding $A$ from $C$;  the condition for regions $A$, $B$, and $C$ that span only a subset of the vertices follow from strong subadditivity. Equation \eqref{eq:chainH} provides a general characterization of such states, which motivates the following definition.
\begin{definition}[Shield commuting Hamiltonian]
Given a graph $G = (V,E)$ and a Hamiltonian $H$ acting  on the particles located in $V$, we say that $H$ is shield commuting when it can be written as $H = H_{AB}+H_{BC}$ with $[H_{AB},H_{BC}] = 0$ whenever $B$ shields $A$ from $C$.
\end{definition}
Thus, \eq{chainH} implies a complete equivalence between positive quantum Markov networks and Gibbs states arising from shield commuting Hamiltonians. Clearly, the local commuting Hamiltonians form a subset of shield commuting Hamiltonians. In particular, it can be easily verified that the terms of a local Hamiltonian commute, but the shield commutation condition involves many-body operators, and there is no obvious efficient way to test it. 

It is the purpose of the this section to examine whether shield commuting Hamiltonians are equivalent to local commuting Hamiltonians. In 1D it follows from the decomposition given in  Eq. \ref{thm:LP} that a positive Markov state $\log{\rho}=H$ may be decomposed as  $H=\sum_i h_{ii+1}$. Equation~\eqref{eq:chainH} then implies that for each pair of terms $[h_{i-1i},h_{ii+1}]=0$.  Such an equivalence has also been shown for cycle-free graphs in \cite{PH11a}, and in Theorem \ref{thm:2cliques} we will extend it to all graphs containing no triangles. For more general lattices however, it is not always the case that a shield commuting Hamiltonian is a local commuting Hamiltonian, as will be demonstrated by the following example.

\begin{figure}
\includegraphics[width=8cm]{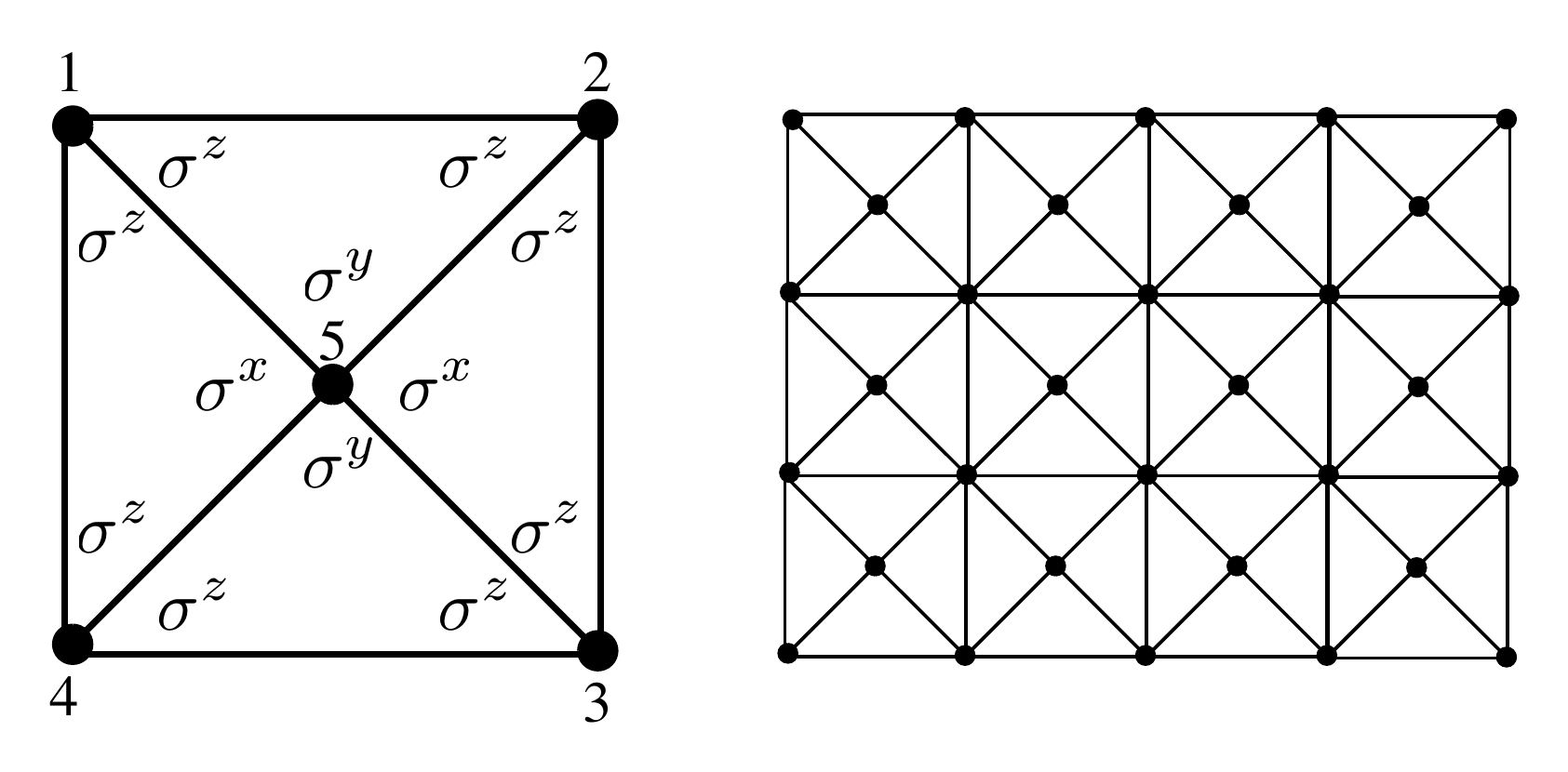}
\caption{Left: Local Hamiltonian on a 5-vertex graph. Each edge in the graph is a clique, but the associated Hamiltonian terms $h_Q=0$ are null. The four triangular cells in the graph are three-body cliques, and the associated Hamiltonian terms are $h_\bigtriangledown = \sigma^z_1\sigma^z_2\sigma^y_5$, $h_{\lhd} = \sigma^z_2\sigma^z_3\sigma^x_5$, $h_\triangle = \sigma^z_3\sigma^z_4\sigma^y_5$, $h_\rhd = \sigma^z_4\sigma^z_1\sigma^x_5$. Right: Lattice tiling with the graph on the left.}
\label{fig:counter_example}
\end{figure}

We now establish that there exist positive quantum Markov networks that are not Gibbs states of local commuting Hamiltonians. Consider the graph $G$ illustrated on the left of Fig.~\ref{fig:counter_example} and its associated Hamiltonian $H = h_\bigtriangledown + h_\lhd + h_\triangle + h_\rhd$. There are only two choices of regions $A$, $B$, and $C$ for this graph such that $B$ shields $A$ from $C$, namely $A = \{1\}$, $B = \{2,4,5\}$, $C = \{3\}$; and $A = \{2\}$, $B = \{1,3,5\}$, $C = \{4\}$. For the first choice, we have $H_{AB} = h_\bigtriangledown + h_\rhd$ and $H_{BC} = h_\lhd + h_\triangle$, and we can easily verify that $[H_{AB},H_{BC}]=0$, which ensures that $I(A:C|B) = 0$. The second choice has $H_{AB} = h_\bigtriangledown + h_\lhd$ and $H_{BC} = h_\rhd + h_\triangle$, and leads to the same conclusion. Thus, the pair $(\rho = \frac 1Z e^H,G)$ is a positive quantum Markov network. Nevertheless, $\rho$ is not the Gibbs state of a local commuting Hamiltonian as the individual terms in the Hamiltonian do not mutually commute.

The choice of the Hamiltonian leading to this situation is very contrived, and the presence of triangular cliques is crucial in our construction.  In fact, we can rule out the existence of such examples in the absence of triangular cliques.
\begin{theorem}
Let  $G = (V,E)$ be a graph and $\rho$ be a density matrix for the particles located at the vertices of $G$. If the pair $(\rho,G)$ is a positive quantum Markov network and $G$ contains only two-body cliques, then the state $\rho$ can be expressed as $\rho = e^H$ where
\begin{equation}
H = \sum_{Q\in \cC} h_Q,\quad [h_Q,h_{Q'}] = 0
\label{eq:local_commute}
\end{equation}
is the sum of mutually commuting Hermitian operators $h_Q$ on the particles located in cliques $Q$.
\label{thm:2cliques}
\end{theorem}

This theorem can be seen as a generalization of a Theorem presented in the supplementary material of \cite{PH11a}, which is restricted to cycle-free graphs.
\begin{theorem}[Poulin and Hastings \cite{PH11a}]
Let  $G = (V,E)$ be a tree graph and $\rho$ be a density matrix for the particles located at the vertices of $G$. The pair $(\rho,G)$ is a positive quantum Markov network if and only if the state $\rho$ can be expressed as $\rho = e^H$ where
\begin{equation}
H = \sum_{Q\in \cC} h_Q,\quad [h_Q,h_{Q'}] = 0
\label{eq:local_commute}
\end{equation}
is the sum of mutually commuting Hermitian operators $h_Q$ on the particles located in cliques $Q$.
\label{thm:PH}
\end{theorem}

The ``if'' implication is a corollary of Theorem \ref{thm:commute}. To prove the ``only if'' direction of Theorem~\ref{thm:2cliques}, we first need to introduce the concept of a genuine $k$-body operator, closely related to the cumulant expansion. Let $H_X$ be an operator on some set of particles $X = \{1,2,\ldots ,k\}$. We say that $H_X$ is a genuine operator on $X$ if $\Tr_Y H = 0$ for all non-empty $Y \subseteq X$. This implies that the cumulant expansion of $H_X$ contains only one term, $H_X$ itself. Equivalently, we can define a cumulant expansion as the decomposition of an operator into a sum of genuine operators on all subsets of particles. One useful fact about genuine operators is given by the following lemma.

\begin{lemma}
Let $H_{AB}$ be a genuine operator on $AB$ and $H_{BC}$ be a genuine operator on $BC$. Then $[H_{AB},H_{BC}]$ is either 0 or a genuine operator on $AB'C$ where $B'\subseteq B$ and $B'\neq\emptyset$.
\label{lemma:supp}
\end{lemma}

This lemma is proven by considering the operator-Schmidt decomposition of the two operators
\begin{align}
H_{AB} &= \sum_j F_A^j\otimes G^j_B \\
H_{BC} &= \sum_k R_B^k\otimes S^j_C
\end{align}
where $\{F_A^j\},\ \{G_B^j\},\ \{R_B^j\},$ and $\{S_C^j\}$ are sets of orthogonal operators. Then,
\begin{equation}
[H_{AB},H_{BC}] = \sum_{jk} F_A^j\otimes [G^j_B,R_B^k]\otimes S^j_C.
\end{equation}
The lemma follows from the fact that the commutator $[G^j_B,R_B^k] $, if non-zero, cannot be proportional to the identity for finite dimensions.

We now turn to the proof of Theorem \ref{thm:2cliques}. In the case of a graph with only two-body cliques, Theorem \ref{thm:LP} already establishes that the cumulant expansion of the Hamiltonian is a sum of terms on edges and vertices, $H = \sum_{e\in E} K_e + \sum_{v \in V} K_v$. We will prove Theorem \ref{thm:2cliques} by expressing the Hamiltonian as $H = \sum_{e\in E} h_e + \sum_{v\in V} h_v$ where all terms mutually commute.  Note that the terms $h_e$ and $h_v$ can differ from the cumulants $K_e$ and $K_v$. In particular, a one-body cumulant $K_v$ can be split arbitrarily among $h_v$ and the edge terms $h_e$ having vertex $v$ as an endpoint, so $h_e$ needs not be a genuine operator on the edge $e$.

We first prove that all two-body cumulants $K_e$ mutually commute. Pick any disjoint regions $A$, $B$, and $C = V-A-B$, such that $B$ shields $A$ from $C$, so we can write $H = {H_{AB}+H_{BC}}$ with $[H_{AB},H_{BC}] = 0$, c.f. Eqs.~(\ref{eq:chain}, \ref{eq:chainH}). We can express $H_{AB}$ and $H_{BC}$ in terms of the cumulants of $H$, with some cumulants contributing to $H_{AB}$ and some to $H_{BC}$. The contribution of most cumulants is unambiguous, but the ones supported only on region $B$ could be either attributed to $H_{AB}$ or $H_{BC}$. More generally, the cumulant expansions have the form
\begin{align}
H_{AB} &= \sum_{ab} K_{ab} + \sum_a K_{a} +\sum_{aa'} K_{aa'} + \sum_{X\subset B} K^A_X \\
H_{BC} &= \sum_{bc} K_{bc} +  \sum_{c} K_c  + \sum_{cc'} K_{cc'}  + \sum_{X\subset B} K_X^C
\end{align}
where
\begin{equation}
 \sum_{X\subset B} K^A_X +   K_X^C  = \sum_{bb'\in B} K_{bb'} +  \sum_{b} K_b.
 \label{eq:sums}
 \end{equation}
and where it is understood that small case letters run only over the vertices in the set labeled with the same capital letter, i.e. $a\in A$, $b\in B$, and $c\in C$. In these equations, we use a double index such as $ab$ to denote an edge $e=(a,b)$ and set $K_{ab} = 0$ if $(a,b) \notin E$. In those terms, the commutation relation $[H_{AB},H_{BC}] = 0$ becomes
\begin{align*}
0 &= \sum_{abc} [K_{ab},K_{bc}] \\
&+ \sum_{bc,X\subset B} [K^A_X,K_{bc}] + \sum_{ab,X\subset B} [K_{ab},K^C_X] \\
&+\sum_{X,Y \subset B} [K^A_X,K_Y^C]
\end{align*}
Note that each summand in the first line has a support that differs from all other terms in the equation.
By Lemma~\ref{lemma:supp}, they must be individually equal to 0. By varying over all possible choices of regions $A$, $B$, and $C$, the first line shows that the two-body cumulants mutually commute, as claimed.

Now that we have established that the two-body cumulants $K_e$ mutually commute, our goal is to split each single-body cumulant $K_u$ among the edge terms $h_e$ acting on site $u$ and the vertex term $h_u$ to obtain mutually commuting terms. More formally, let $\cN(u) = \{v\in V:(u,v)\in E\}$ denote the neighborhood of $u$, and denote the degree of $u$ by $d(u) = |\cN(u)|$. We will decompose  $K_u = h_u + \sum_{v\in \cN(u)} G_u^{v}$, and define the edge terms
\begin{equation}
h_{uw}  = K_{uw} +  G^u_w + G_u^w
\end{equation}
in a way that $[h_u,h_{uv}] = 0$ and $[h_{uw},h_{uv}] = 0$, or equivalently
\begin{align}
[h_{uw},h_{uv}] &= [K_{uw} +  G^u_w + G_u^w,K_{uv} +  G^u_v + G_u^v] \\
&=  [K_{uw} ,G_u^v] +  [G_u^w,K_{uv}] + [G_u^v,G_u^w] = 0.
\end{align}
Since the three commutators in the last line have distinct support, by Lemma~\ref{lemma:supp} this last condition is equivalent to  $[K_{uw} ,G_u^v] =0$ and $[G_u^v,G_u^w] =0$ for all $(u,w)\in E$ and $(u,v) \in E$. What remains to be explained is how to choose the $G_u^v$.

\begin{figure}
\includegraphics[width=5cm]{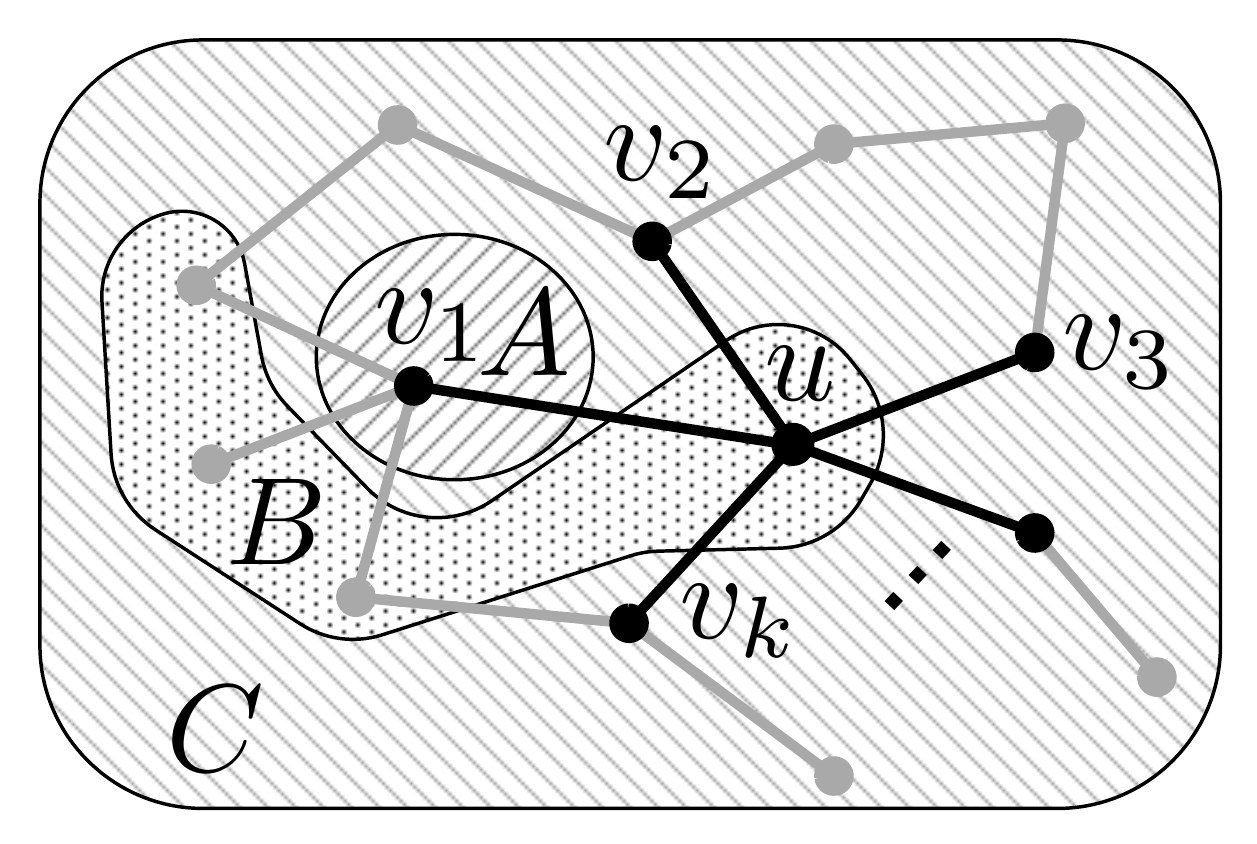}
\caption{Definition of the regions $A$, $B$, and $C$ for a given neighbor $v_1$ of vertex $u$.}
\label{fig:neighbors}
\end{figure}

Consider a vertex $u$ in the graph. Let $\{v_1,v_2,\ldots v_k\} = \cN(u)$ denote the immediate neighbors of $u$. Define three sets of vertices
\begin{align}
A & \subset \cN(u)\\
B &= \cN(A) \\
C &= V-A-B.
\end{align}
where the neighbors of the set $A$ are defined in the natural way $\cN(A) = \{w\in V: \exists v \in A, (w,v) \in E\}$. These definitions are illustrated on Fig.~\ref{fig:neighbors}. In other words, $B$ consists of the neighbors of a subset $A$ of the neighbors of $u$, so in particular $B$ includes the vertex $u$ itself. It follows that $B$ shields $A$ from $C$. Moreover, because the graph has only two-body cliques, all the neighbors $v_l\notin A$ are contained in $C$. The Markov condition implies that $H = {H_{AB}+H_{BC}}$ with $[H_{AB},H_{BC}] = 0$ for this choice of regions, c.f. Eqs.~(\ref{eq:chain}, \ref{eq:chainH}).

Consider the cumulant expansion of $H_{AB}$ and $H_{BC}$. Because $H = H_{AB} +H_{BC}$, the cumulants in these expansions can only differ from those of $H$ on the region $B$, i.e. we must have
\begin{align}
H_{AB} &= \sum_{ab} K_{ab} + \sum_a K_a + \sum_{X\subset B} K_X^A \label{eq:cumA}\\
H_{BC} &= \sum_{bc} K_{bc} +  \sum_c K_c  + \sum_{X\subset B} K_X^C + \sum_{cc'} K_{cc'}
\end{align}
where
\begin{equation}
 \sum_{X\subset B} K^A_X +  \sum_{X\subset B} K_X^C  = \sum_{bb'\in B-u} K_{bb'} +  \sum_{b} K_b.
 \label{eq:sums}
 \end{equation}
Note that \eq{cumA} cannot contain two-body cumulants $K_{aa'}$ because this would create a triangle in the graph, which is ruled out in the hypothesis of the Theorem. For the same reason, the two-body cumulants $K_{bb'}$ in \eq{sums} cannot include node $u$. Equation~\ref{eq:sums} implies that $K_X^A = -K_X^C$ for all $|X|>2$ as well as for all $|X| = 2$ when $u\in X$. Now, consider the commutation $[H_{AB},H_{BC}] = 0$ in this expansion:
\begin{align}
0 &= \sum_{abc}[K_{ab},K_{bc}] \label{eq:comm1}\\
&+ \sum_{ab X} [K_{ab},K_X^C] +\sum_{bcX} [K^A_X,K_{bc}] \\
&+ \sum_{XY} [K_X^A,K_Y^C].
\end{align}
The first line \eq{comm1} is 0 since two-body cumulants of $H$ mutually commute. By Lemma~\ref{lemma:supp}, each individual term in the second line must be 0 because they each have a distinct support from all other non-zero terms in the sum.  In the last line, by Lemma~\ref{lemma:supp}, the only terms in the sum that can be supported on the single site $u$ are of the form $[K^A_u,K_u^C]$ or  $[K^A_X,K^C_X]$ with $u\in X$ and $|X|>1$, but these latter are 0 since $K_X^A = -K_X^C$ as shown above.  Therefore, the terms with $X = Y = u$ are decoupled from the other terms, and so must obey $[K^A_u,K_u^C] = 0$. We conclude that for any subset $A\subset \cN(u)$, we can express $K_u = K_u^A + K_u^C$ with
\begin{equation}
[K_u^A,K_u^C] = [K_{au},K_u^C] = [K_u^A,K_{uc}] = 0
\label{eq:comm}
\end{equation}
for all $a\in A$, and $c\in \cN(u)-A$.

Consider the subgraph $G(u)$ of $G$ consisting of site $u$ and its neighbors $v_j$,  and define the Hamiltonian $H(u) = \sum_v K_{uv} + K_u$. Note that $G(u)$ is a tree and that the pair $(\rho(u) = \frac 1Z e^{H(u)},G(u))$ is a positive quantum Markov network. This last fact follows from the commutation relations of \eq{comm} and Theorem~\ref{thm:commute}. By Theorem~\ref{thm:PH}, it follows that $K_u$ can be decomposed into $K_u =  h_u + \sum_{v\in \cN(u)} G_u^v$  in such a way that $h'_{uv} = K_{uv} + G_u^v$ and $h_u$ all mutually commute. Repeating the argument for every node $u\in V$ and defining $h_{uv} =  K_{uv} + G_u^v + G_v^u$ complete the proof of Theorem~\ref{thm:2cliques}.

Note that the proof of Theorem~\ref{thm:2cliques} makes crucial use of the fact that the lattice has only two-body cliques. As a consequence, the proof does not extend directly to Hamiltonians that are the sums of two-body terms but embedded on, say, a triangular lattice.

\section{Conclusion and Discussion}

In this article, we have presented a simplified proof that all positive quantum Markov networks living on lattices of finite dimensional quantum systems are Gibbs states of Hamiltonians local to the cliques of the corresponding graph; proved that there is an equivalence between Gibbs states of commuting Hamiltonians and positive quantum Markov networks on graphs that do not contains triangles, shown that Gibbs states arising from local Hamiltonians with commuting terms are positive quantum Markov networks; and lastly; demonstrated that there exist positive quantum Markov networks on triangular lattices that cannot be written as Gibbs states of commuting local Hamiltonians.  While both classical and quantum positive Markov networks are Gibbs states of Hamiltonians local to the cliques of their underlying graph, the essential difference is that the terms of a classical Hamiltonian commute by construction, whereas the terms in the quantum Hamiltonians are only required to commute when grouped into globally defined regions as in Fig.\ref{fig:ABC}. 

It remains an open question whether there exist Hamiltonians that satisfy this global commutation property---that we called shield commutation---but that cannot be transformed into a local commuting Hamiltonian under a suitable renormalization procedure. Despite many attempts, we have been unable to construct such models. This is illustrated by the example of Sec.~\ref{sec:comm}.  If the network is extended to a regular 2D lattice as shown on the left of Fig.~\ref{fig:counter_example}, we again obtain a Gibbs state that is a quantum Markov network, yet the Hamiltonian terms do not commute. However, if we coarse grain the lattice by combining the central spin of each unit cell to the spin immediately to its northeast---and thus join $ h_\bigtriangledown$ and $h_\rhd$ and similarly  $h_\bigtriangleup$ and $h_\lhd$  to the same cliques---we obtain a local commuting Hamiltonian. This illustrates that the lattice model can be fixed by a local rearrangement of the degrees of freedom, but has no ``large-scale" obstructions to commutation. 

It has been established \cite{AE11a,BV05a} that for Hamiltonians containing only two-body commuting interactions, the Hilbert space of each vertex must split into a direct sum of factor spaces as in Eq. \ref{eq:split}. This is reminiscent to the fact that we can only prove an equivalence between local commuting Hamiltonians and positive quantum Markov networks in the absence of triangular cliques. Although the local splitting property is not necessary for a factorization into commuting terms---the projector into the code space of Kiteav's toric code \cite{Kit03a} is a quantum Markov network which factorizes by construction---, it is sufficient. Since commutation of genuine operators that act non-trivially on more than two subsystems does not imply that the local Hilbert spaces will split \cite{BV05a}, quantum Markov networks containing triangles and hence three-body cliques may involve complex non-local structures, which could allow for the shield and local commuting properties to be inequivalent even under coarse graining.

A folk theorem in condensed matter physics states that all phases of matter can be realized with local commuting Hamiltonians.  This is supported by the fact that Markov networks are fixed points of a renormalization procedure. A corollary of this statement would be that, under a suitable renormalization procedure, all Gibbs states are quantum Markov networks, thus establishing a complete equivalence between Gibbs states and quantum Markov networks. If models with large-scale obstructions to commutation could be found on the other hand---Hamiltonians that are shield commuting but cannot be locally transformed into local commuting Hamiltonians---, they would reveal a new phase of matter that exhibits quantum non-locality without long range entanglement \cite{BDF+99a}, and would require a refinement of this folk theorem. They would also  form a class of Hamiltonians of intermediate complexity between commuting and general local Hamiltonians \cite{K02a,BV05a,S11a,AE11a}.

Lastly, the relation between the Markov condition and area laws is also intriguing. At finite temperature, the entropy of a region should be mostly extensive---scaling with its volume---but the zero-temperature area-law should translate in an additional boundary contribution. Thus, we can expect a generic scaling $S(A) = \alpha |A| + \beta |\partial A|$ for large enough regions $A$. A simple geometric argument shows that under this scaling, regions $A$, $B$, $C$ chosen as in Fig.~\ref{fig:ABC} always obey the Markov condition.  This gives us additional reasons to believe that Markov networks are generic properties of Gibbs states on sufficiently coarse grained networks.

\section{Acknowledgements}

We thank Andy Ferris, Matt Hastings, Olivier Landon-Cardinal,  Matt Leifer, and John Preskill for stimulating discussions. This work was partially funded by NSERC, CRM, Mprime, and the Lockheed Martin Coorporation.


\end{document}